\documentclass[12pt,fleqn]{article}
\usepackage[textheight=23cm,textwidth=16cm]{geometry}
\usepackage[utf8]{inputenc}
\usepackage{amsmath}
\usepackage{graphicx}
\usepackage[superscript]{cite}
\usepackage[]{hyperref}
\usepackage[uncertainty-mode=compact, uncertainty-mode=separate, list-units=single, range-units=single]{siunitx}
\usepackage[version=4]{mhchem} 

\usepackage{xcolor}

\usepackage[american]{babel}

\usepackage{dcolumn} 
\newcolumntype{d}{D{.}{.}{2}}
\newcolumntype{e}{D{.}{.}{3}}
\newcolumntype{f}{D{.}{.}{4}}

\def\boldr{{\mathbf{r}}}

\begin{document}

\begin{center}

{\LARGE\bf
  Interoperable Workflows by Exchanging Grid-Based Data between Quantum-Chemical Program Packages \\
}

\vspace{1cm}

{\large 
Kevin Focke$^{a,}$\footnote{ORCID: 0009-0009-9079-1769},
Matteo De Santis$^{b,}$\footnote{ORCID: 0000-0001-7366-1780},
Mario Wolter$^{a,}$\footnote{ORCID: 0000-0001-6948-6801},
Jessica A. \\ Martinez B$^{b,c,}$\footnote{ORCID: 0000-0002-9577-3449},
Valérie Vallet$^{b,}$\footnote{ORCID: 0000-0002-2202-3858},
André Severo Pereira Gomes$^{\ast,b,}$\footnote{ORCID: 0000-0002-5437-2251, E-Mail: andre.gomes@univ-lille.fr},
Małgorzata Olejniczak$^{\ast,d,}$\footnote{ORCID: 0000-0002-8370-9570, E-Mail: malgorzata.olejniczak@cent.uw.edu.pl},
Christoph R. Jacob$^{\ast,a,}$\footnote{ORCID: 0000-0002-6227-8476, E-Mail: c.jacob@tu-braunschweig.de}
}\\[4ex]

{\small \linespread{1.1}\selectfont
$^{a}$ Technische Universität Braunschweig, Institute of Physical and Theoretical Chemistry, \\
Gaußstraße~17, 38106 Braunschweig, Germany \\[1ex]

$^{b}$ Univ. Lille, CNRS, UMR 8523 – PhLAM – Physique des Lasers Atomes et Molécules, F-59000 Lille, France \\[1ex]

$^{c}$ Department of Chemistry, Rutgers University, Newark, New Jersey, USA\\[1ex]

$^{d}$ Centre of New Technologies, University of Warsaw, S.\ Banacha 2c, 02-097 Warsaw, Poland\\
}

\vspace{18ex}

\end{center}

\begin{tabbing}
Date:   \qquad\qquad\quad \= March 29, 2024 \\
\end{tabbing}

\newpage

\begin{abstract}

Quantum-chemical subsystem and embedding methods require complex workflows that may involve 
multiple quantum-chemical program packages. Moreover, such workflows require the exchange of voluminous
data that goes beyond simple quantities such as molecular structures and energies. Here, we describe
our approach for addressing this interoperability challenge by exchanging electron densities and 
embedding potentials as grid-based data. We describe the approach that we have implemented 
to this end in a dedicated code, \textsc{PyEmbed}, currently part of a Python scripting framework. We discuss how it has 
facilitated the development of quantum-chemical  subsystem and embedding methods, and highlight 
several applications that have been enabled by \textsc{PyEmbed}, including WFT-in-DFT embedding 
schemes mixing non-relativistic and relativistic electronic structure methods, real-time time-dependent DFT-in-DFT approaches, the density-based many-body expansion, 
and workflows including real-space data analysis and visualization. Our approach demonstrates in particular the merits of exchanging (complex) grid-based data, and in general the 
potential of modular software development in quantum chemistry, which hinges upon libraries that
facilitate interoperability.

\end{abstract}

\newpage

\section{Introduction}

Contemporary applications of quantum chemistry increasingly rely on complex workflows, which combine calculations with 
different methods and program packages. For instance, schemes for the automatic exploration of chemical reaction networks 
(see, e.g., Ref.~\citenum{dewyer_methods_2018, unsleber_exploration_2020, baiardi_expansive_2022}) combine heuristics 
for the generation of possible reaction products and pathways \cite{rappoport_complex_2014, bergeler_heuristics-guided_2015, 
rappoport_predicting_2019}, semi-empirical quantum chemistry for the generation of conformer ensembles \cite{bannwarth_gfn2-xtb_2019, 
pracht_automated_2020}, and quantum-chemical calculations at different levels of theory, possibly in combination with approaches 
for quantifying their uncertainty \cite{simm_error-controlled_2018, proppe_mechanism_2019}. Similarly, the elucidation of reaction 
mechanisms in homogeneous transition-metal catalysis requires intricate workflows combining distinct computational 
tools (for examples, see Ref.~\citenum{podewitz_origin_2021, talmazan_encapsulation_2022}). Complex computational workflows 
are further at the heart of high-throughput applications of quantum chemistry \cite{jain_commentary:_2013, hachmann_lead_2014, 
pyzer-knapp_what_2015, unsleber_high-throughput_2023}, which are often a prerequisite for leveraging the potential of machine 
learning in chemistry \cite{kulik_making_2020, pollice_data-driven_2021, nandy_computational_2021, rankine_progress_2021, 
kirschbaum_machine_2023, eckhoff_quantitative_2024}. Numerous software packages have been developed over the past two
decades to facilitate such computational workflows in quantum chemistry \cite{pyadf-2011, yang_jacob:_2012, jain_fireworks:_2015,
arshad_multi-level_2016, larsen_atomic_2017, guan_aaron:_2018, zapata_qmflows:_2019, huber_aiida_2020, uhrin_workflows_2021,
george_automation_2021, ingman_qchasm:_2021, smith_quantum_2021, unsleber_chemoton_2022, scine, hicks_massively_2023, 
dral_mlatom_2023, rosen_jobflow:_2024}.

The computational treatment of complex chemical systems such as biomolecules, molecular aggregates, or nanoscopic materials 
and interfaces is nowadays enabled by multiscale and multilevel methods of computational chemistry \cite{gomes_quantum-chemical_2012,
gordon_fragmentation:_2017, jones_embedding_2020,jansen_quantum-derived_2022}. Examples of such methods 
include QM/MM approaches \cite{senn_qm/mm_2009, sousa_application_2017, magalhes_modelling_2020}, quantum-chemical 
fragmentation methods \cite{raghavachari_accurate_2015,herbert_fantasy_2019, liu_recent_2023}, ONIOM-type 
methods \cite{chung_oniom_2015, seeber_growing_2023}, and density-based subsystem and embedding 
methods \cite{jacob_subsystem_2014, krishtal_subsystem_2015, wesolowski_frozen-density_2015, lee_projection-based_2019,
jacob_subsystem_2024}. 
As these multiscale and multilevel methods combine rather distinct computational approaches for different parts or fragments of 
complex chemical systems, they commonly give rise to computational workflows that combine many individual calculations, 
possibly with different quantum-chemical software modules or packages --- both for density-functional theory (DFT) and for
various methods of wave-function theory (WFT) --- and with other software tools, such as force-field codes or chemoinformatics
toolkits. Furthermore, they might require exchanging data between these individual calculations beyond ``simple'' quantities such 
as atomic coordinates or total energies. Several computational tools for realizing such multiscale workflows have been developed 
in the past decades for different use cases, in particular for QM/MM schemes \cite{metz_chemshell_2014, weingart_cobramm_2018, 
chemshell-2, zhang_janus:_2019, gtze_user-friendly_2021, mart_qmcube_2021, chemshell-3} and for fragmentation 
methods \cite{woodcock_mscale:_2011, odinokov_pyefp:_2018, broderick_scalable_2023}.

To realize the computational workflows that are required for density-based subsystem and embedding methods, some of us 
have initially developed the Python-based scripting framework \textsc{PyAdf} \cite{pyadf-2011}. While initially focused
around the Amsterdam Density Functional (\textsc{Adf}) program \cite{chem-with-adf} [now part of the Amsterdam Modeling
Suite (\textsc{Ams})], over the years \textsc{PyAdf} has been extended to facilitate density-based subsystem and embedding methods 
in which different quantum-chemical methods and program packages can be combined. 

In this context, the interoperability of different quantum-chemical program packages poses particular challenges. 
Commonly, quantum-chemical program packages (such as \textsc{Ams}\cite{chem-with-adf, ams-2021}, \textsc{Dalton} \cite{aidas_dalton_2014,
olsen_dalton_2020}, \textsc{Dirac}\cite{saue_dirac_2020}, \textsc{NWChem}\cite{dam_nwchem:_2011, apr_nwchem:_2020}, 
\textsc{Molcas}\cite{aquilante_modern_2020, li_manni_openmolcas_2023}, \textsc{Orca}\cite{neese_orca_2020, neese_software_2022}, 
\textsc{Turbomole}\cite{turbomole-jcp-2020, franzke_turbomole:_2023}, and \textsc{QuantumEspresso}\cite{giannozzi_quantum_2020,
carnimeo_quantum_2023}) have been developed as monolithic codes over
many decades\cite{jacob_open_2016}. Therefore, there is generally no easy access to the internals of the program packages, 
either because they are closed-source codes or because they are too complex for users and developers outside of their core 
community. However, density-based subsystem and embedding methods require exchanging information between different
program packages and/or the workflow environment that go beyond those commonly provided to the user, such as integration
grids or electron densities and derived properties. Unfortunately, there are no standardized interfaces for exchanging such information between different quantum-chemical program packages.

Here, we discuss how these challenges have been addressed in the \textsc{PyAdf} scripting framework. To this end, we describe 
the \textsc{PyEmbed} module of \textsc{PyAdf} for handling grid-based data (see Section~\ref{sec:design}) and demonstrate how 
it facilitates interoperable workflows in multiscale quantum chemistry.
To illustrate the capabilities of our approach, we discuss selected examples of workflows that either combine different program 
packages or that are agnostic to the underlying quantum-chemical programs and methods. In particular, we highlight the 
development and applications of DFT-in-DFT and WFT-in-DFT embedding schemes (see Section~\ref{sec:wftindft}), of 
real-time time-dependent DFT-in-DFT embedding schemes (see Section~\ref{sec:rt}), and of the density-based many-body
expansion (see Section~\ref{sec:dbmbe}). Furthermore, we discuss the integration of real-space data analysis and 
visualization into computational workflows (see Section~\ref{sec:gosia-ttk}). 
Note that the present article focuses on \textsc{PyAdf}'s functionality for handling grid-based data and does not aim at
covering its full capabilities in other areas, which will be discussed elsewhere.

\section{Challenges and General Design}
\label{sec:design}

The scripting framework \textsc{PyAdf} allows the user to set up workflows combining several quantum-chemical calculations seamlessly.
It provides classes for handling and manipulating atomic coordinates, for defining quantum-chemical calculations with
a number of program packages (including \textsc{Adf}/\textsc{Ams}, \textsc{Dalton}, \textsc{Dirac}, \textsc{NWChem}, \textsc{Molcas}, 
\textsc{Orca}, and \textsc{Turbomole}) as well as for
generating input files for these programs, for executing the quantum-chemical calculations as well as for storing their results,
and for extracting results data (such as optimized structures, energies, and properties) from the result files. The general
design of \textsc{PyAdf} is described in Ref.~\citenum{pyadf-2011}, even though its functionality has been extended significantly in
the past decade.

\subsection{Use Case: Density-Based Embedding}

The prototypical use case that is of interest to us here are density-based subsystem and embedding calculations, in
particular, DFT-in-DFT and WFT-in-DFT calculations based on the framework of subsystem-DFT and frozen-density 
embedding (FDE)  \cite{jacob_subsystem_2014, wesolowski_frozen-density_2015, jacob_subsystem_2024}. Here,
one considers a molecular system, which is divided into $N$ subsystems. This is achieved by partitioning the total 
electron density $\rho_\text{tot}(\boldr)$ into the electron densities of these subsystems,
\begin{equation}
  \rho_\text{tot}(\boldr) = \sum_I \rho_I(\boldr).
\end{equation}
When considering the first subsystem as the subsystem of interest, its electron density $\rho_1(\boldr)$ is obtained
by minimizing the total energy functional. This results in Kohn-Sham (KS)-like equations for the active subsystem,
in which the effect of the frozen environment of all the other subsystems (with electron densities $\rho_{J=2,\dotsc,N}$) 
is included via an effective embedding potential \cite{tomasz-1993},
\begin{align}
\label{eq:embpot}
v^{(1)}_{\rm emb}[\rho_1, \rho_{\rm tot}](\boldr) 
   =& \sum_{I=2}^N v_{\rm nuc}^{(I)}(\boldr) + v_{\rm Coul}[\rho_{\rm tot} - \rho_I](\boldr) \nonumber \\
        &\qquad +\ \frac{\delta E_\text{xc}[\rho_{\rm tot}]}{\delta\rho_{\rm tot}(\boldsymbol{r})} - \frac{\delta E_\text{xc}[\rho_1]}{\delta\rho_1(\boldsymbol{r})}
        + \frac{\delta T_s[\rho_{\rm tot}]}{\delta\rho_{\rm tot}(\boldsymbol{r})} - \frac{\delta T_s[\rho_1]}{\delta\rho_1(\boldsymbol{r})}.
\end{align}
Here, $v_{\rm nuc}^{(I)}(\boldr)$ is the potential of the nuclei in subsystem~$I$, $v_{\rm Coul}[\rho](\boldr)$ is the
Coulomb potential of electron density~$\rho$, $E_\text{xc}[\rho]$ is the exchange--correlation functional, and
$T_s[\rho]$ is the kinetic energy functional. The latter two are commonly evaluated using suitable approximate 
density functionals.

For DFT-in-DFT embedding calculations, this embedding potential is included in the KS equations. Note that it
depends on the electron density of the active subsystem, i.e., it needs to be updated during the self-consistent
field interactions. In WFT-in-DFT calculations, the embedding potential is included in the one-electron Hamiltonian,
which requires evaluating the matrix elements,
\begin{equation}
  \label{eq:ao-emb-ints}
  V^{\text{emb}}_{\mu\nu} 
    = \langle \chi_\mu | v_{\rm emb} | \chi_\nu \rangle
    = \int  \chi^*_\mu(\boldr) v_{\rm emb}(\boldr) \chi_\nu(\boldr) \, {\rm d}^3r,
\end{equation}
where $\chi_\mu$ are the basis functions. For further details on the inclusion of the FDE embedding potential
in WFT calculations, see Refs.~\citenum{gomes_calculation_2008, gomes_quantum-chemical_2012, hfener_molecular_2012, 
hfener_calculation_2012, gomes_towards_2013}. Here, one commonly replaces the dependence of the embedding potential 
on the electron density of the active subsystem by a fixed density (linearized embedding potential\cite{duak_linearized_2009}).

Several quantum-chemical program packages provide integrated implementation of DFT-in-DFT embedding, e.g., 
\textsc{Adf}/\textsc{Ams}\cite{newfde-2007}, Serenity\cite{unsleber_serenity:_2018, niemeyer_subsystem_2023}, CP2K\cite{andermatt_combining_2016}, 
and eQE\cite{genova_eqe:_2017, mi_eqe_2021}. These control the calculations for the different subsystems themselves 
and use their internal data structures to exchange data between them. In this case, algorithms and routines available
within these programs can be used to evaluate the different terms of the embedding potential and their matrix elements.
An integrated implementation of WFT-in-DFT methods is available in \textsc{Koala}\cite{hfener_koala_2021}.

\begin{figure}
\centering
  \includegraphics[width=0.65\linewidth]{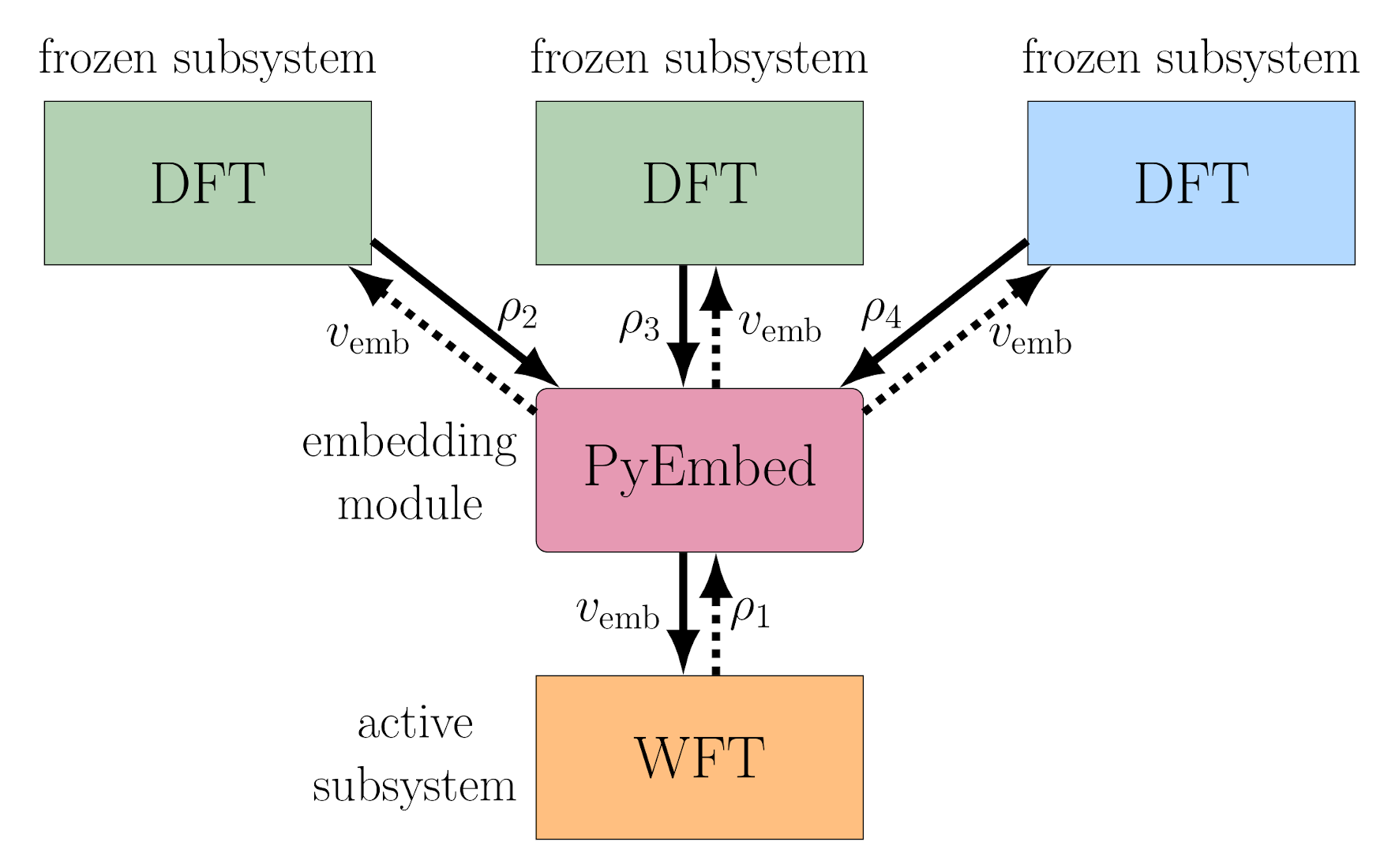}
  \caption{\linespread{1.0}\selectfont
                Illustration of a modular approach to WFT-in-DFT calculations. Each subsystem is treated in a separate 
                quantum-chemical calculation (shown as rectangles), possibly using different program packages (indicated 
                by the colors of the rectangles). The electron densities from the individual calculations serve as input to 
                \textsc{PyEmbed}, which calculates the embedding potential that is subsequently imported into the WFT calculation 
                of the active subsystem. In freeze-and-thaw calculations, the roles of the different subsystems can be 
                interchanged (indicated by the dashed arrows).}
\label{fig:wft-in-dft}
\end{figure}

However, such an integrated approach makes it difficult to combine the unique capabilities of different program packages 
in a single DFT-in-DFT or WFT-in-DFT workflow. An alternative modular approach, which is facilitated by the \textsc{PyEmbed} 
module, is illustrated in Fig.~\ref{fig:wft-in-dft}. Here, DFT calculations for the frozen subsystems are performed individually,
possibly using different methods and/or program packages for different frozen subsystems (see green and blue rectangles
at the top of the figure). These calculations provide the electron densities of the subsystems, from which the
\textsc{PyEmbed} module calculates the embedding potential according to Eq.~\eqref{eq:embpot} (see central box), which is 
subsequently imported into the WFT calculation for the active subsystem, again using a different program package (see 
orange box at the bottom). In freeze-and-thaw calculations, roles of the subsystems can be interchanged, and the electron
density from the WFT calculation (or an approximation thereof) can be used to obtain an embedding potential for the
initially frozen subsystems, leading to updated frozen electron densities that can, in turn, be used to calculate an updated
embedding potential for the WFT calculation.

Such a modular approach to DFT-in-DFT and WFT-in-DFT calculations poses a challenge, as it requires clearly defined
interfaces between the quantum-chemical program packages and \textsc{PyEmbed} that facilitate interoperability. In particular,
the quantum-chemical program packages, on the one hand, need to be able to export the electron density and related 
quantities, and on the other hand, must provide the capability to import an external embedding potential. In particular
the latter will generally require modifying the quantum-chemical program packages themselves, but the required changes
should be as lightweight as possible.

Here, we chose to address these challenges by exchanging the densities and potentials as grid-based data. That is, we
employ a suitable three-dimensional grid of points $\{\boldr_i\}$, and express the electron densities and related quantities
as well as embedding potentials on these grid points. Usually, the grid points coincide with those of a suitable numerical integration
grid (e.g., a Becke grid \cite{becke_multicenter_1988}), which makes it possible to evaluate the integrals over the
embedding potential [Eq.~\eqref{eq:ao-emb-ints}] by numerical integration:
\begin{equation}
  \label{eq:ao-emb-ints-numint}
  V^{\text{emb}}_{\mu\nu} 
    = \langle \chi_\mu | v_{\rm emb} | \chi_\nu \rangle
    \approx \sum_i  \chi^*_\mu(\boldr_i) v_{\rm emb}(\boldr_i) \chi_\nu(\boldr_i).
\end{equation} 
To this end, the embedding potential is also evaluated at the individual grid points from the values of the electron 
densities and related properties at the grid points, which are exported from the quantum-chemical program packages
on these grid points. In addition to the electron density itself, the first and second derivatives of the electron density
at the grid points are required to calculate the functional derivatives of $E_\text{xc}[\rho]$ and of $T_s[\rho]$ when
using generalized-gradient approximations (GGAs), and the Coulomb potential of the electron densities,
\begin{equation}
   v_{\rm Coul}[\rho](\boldr) = \int \frac{\rho(\boldr')}{|\boldr - \boldr'|} \, {\rm d}^3r,
\end{equation}
should also be provided on the grid points by the quantum-chemical program packages. Furthermore, the nuclear
potential $v_{\rm nuc}(\boldr)$ at each grid point is required, but can be easily calculated directly from the atomic 
coordinates.

Exchanging grid-based data facilitates interoperability, as it defines a clear interface that is independent of the 
quantum-chemical methods that are used, of the employed basis functions and the specific basis sets, and of 
the inner workings of the individual quantum-chemical program packages.  In the following, we will present the
primary components of the \textsc{PyEmbed} module that allow for the exchange of grid-based data between quantum-chemical
program packages. 

While our focus here is on calculating embedding potentials using \textsc{PyEmbed}, it is worth noting that exchanging grid-based data can extend to more complex quantities. For instance, it can involve derivatives of the embedding potential, essential for computing kernel contributions to linear response properties \cite{olejniczak_calculation_2017}.

\subsection{Handling of Grid-Based Data}

\begin{figure}
\centering
  \includegraphics[width=0.6\linewidth]{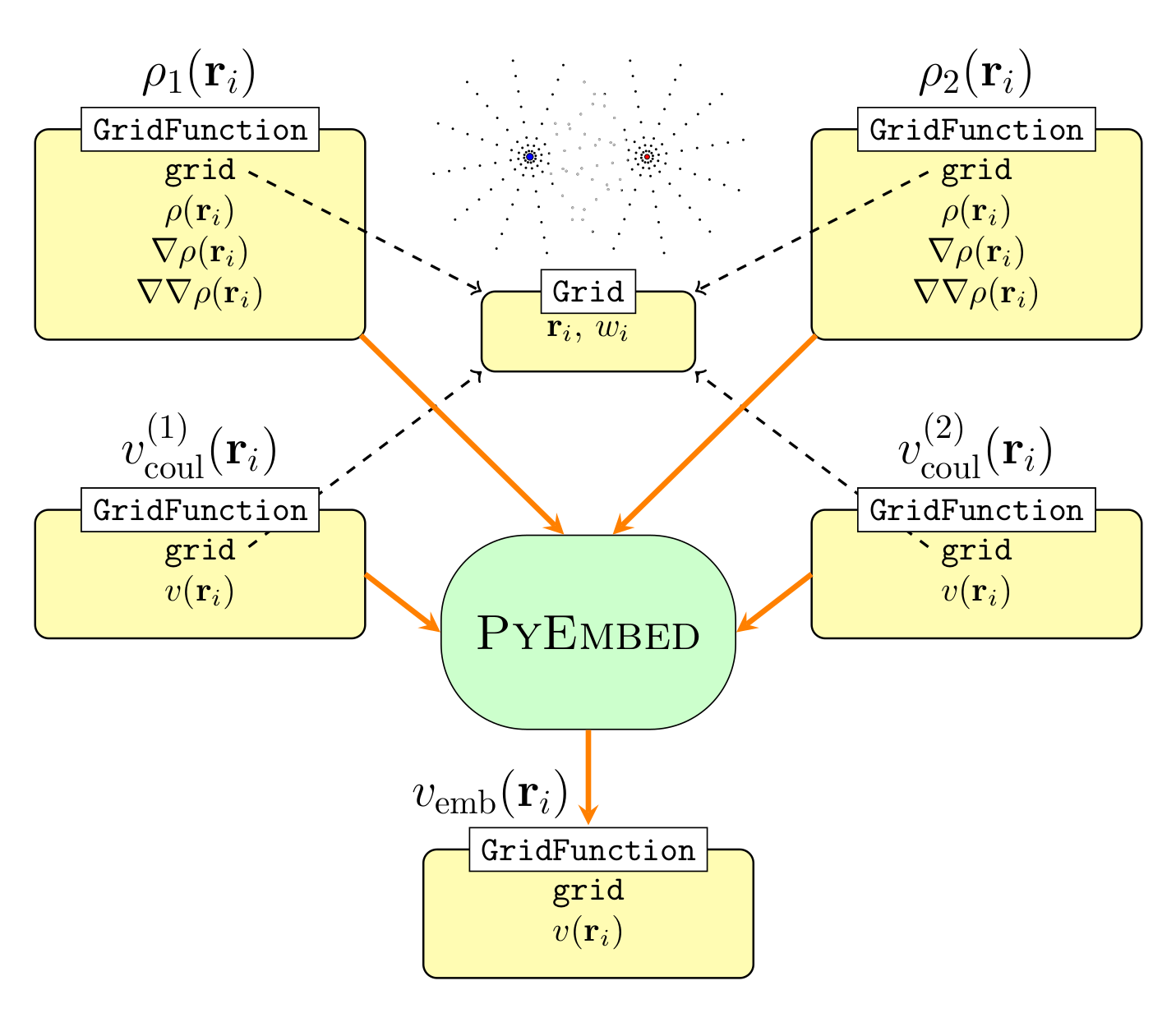}
  \caption{\linespread{1.0}\selectfont
                Overview of the general design of \textsc{PyEmbed}, for the example of the evaluation of the embedding
                potential $v_\text{emb}(\boldsymbol{r}$ from the fragments electron densities $\rho_1(\boldr)$ and 
                $\rho_2(\boldr)$ as well as the corresponding Coulomb potentials $v_\text{Coul}^{(1)}(\boldr)$ and
                $v_\text{Coul}^{(2)}(\boldr)$. The grid-based representations of the densities and potentials are stored
                as instances of \texttt{GridFunction}s, each referencing the associated \texttt{Grid}. See text
                for further details.}
\label{fig:pyembed-classes}
\end{figure}

To facilitate the handling of grid-based quantities in Python, we have developed a library of \texttt{GridFunction} classes, 
which provides the required data structures (see Fig.~\ref{fig:pyembed-classes}). Simply speaking, a \texttt{GridFunction} 
contains references to the associated integration grid and the relevant data on these grid points. 

Here, the integration grid is represented by an instance of a \texttt{Grid} class, which encapsulates the coordinates of the grid 
points as well as their weights. Different types of grids, such as evenly-spaced cubic grids (commonly used for visualization) 
and irregular molecular integration grids (as used in molecular quantum chemistry) are supported. \textsc{PyEmbed} interfaces 
with PySCF \cite{sun_pyscf:_2018, sun_recent_2020} to generate suitable Becke integration grids based on the molecular 
structure, independent of external program packages. The accuracy of the generated Becke grid can be tuned according to
the options available in PySCF.

Grid-based data (values on the grid points) are generally stored in \textsc{Numpy} arrays. Different subclasses 
of \texttt{GridFunction} are tailored to specific common use cases. For instance, \texttt{GridFunction\-DensityWithDerivatives} 
can encapsulate the electron density $\rho(r_i)$ (stored as one-dimensional \textsc{Numpy} array of size $N$), its 
gradient $\boldsymbol{\nabla}\rho(r_i)$ (stored as $N \times 3$ array), and if necessary its second 
derivative $\boldsymbol{\nabla}\boldsymbol{\nabla}\rho(r_i)$ (stored as $N \times 6$ array). The possibility to 
represent spin unrestricted densities is also available.

All \texttt{GridFunction}s provide methods for mathematical operations on them, such as addition and multiplication 
by scalars, as well as for numerical integration. These methods provide an interface that is unified across all types 
of grid functions, and are implemented efficiently via \textsc{Numpy}. Furthermore, all \texttt{GridFunction}s include 
functionality for reading and writing the grid and data to file in various formats, including \textsc{Numpy}'s \cite{numpy, 
numpy-review} text and binary formats, as well as HDF5\cite{hdf5}. Finally, routines for interpolating grid-based
quantities between different grids are also available.

\subsection{Generation of Grid-Based Data}

The available methods for reading \texttt{GridFunction}s from files provide a natural interface to external program packages
and can be used to import electron densities and related quantities, provided the program packages allow for the export
of grid-based data in a suitable file format. Currently, \textsc{PyAdf} supports the import from \textsc{Ams}'s tape files as well as from
text files written by \textsc{Dalton} and \textsc{Dirac}. Implementing interfaces to other program packages and file formats as well as to
other scalar, vector and tensor fields beyond density-related quantities (see Sect.~\ref{sec:gosia-ttk}) is straightforward.

Another approach for obtaining the values of the electron density and related quantities on the grid points proceeds via
the molecular orbital coefficients optimized in the quantum-chemical calculation. Together with information on 
the basis functions used, this allows for the calculation of the electron density and related quantities inside
\textsc{PyEmbed}. However, despite efforts to develop a standardized data format for the exchange of the required basis set 
and orbital / wavefunction information (such as Q5Cost\cite{q5cost}, QCSchema\cite{smith_molssi_2021}, and 
TREXIO\cite{trexio}), only a few quantum-chemical program packages currently provide such a standardized interface. 
Further complications may arise if calculations take into account relativistic effects\cite{dyall-textbook, pyykk_relativistic_2012} 
such as spin-orbit coupling\cite{saue_relativistic_2011}, since then one-electron functions may be expressed as 
complex-valued functions, or as quaternions\cite{saue_quaternion_1999, senjean_generalization_2021}. 
However, if one remains within the non-relativistic theory, the file format used by the \textsc{Molden} 
program \cite{molden} (originally developed several decades ago), provides a \textit{de facto} standard, as such 
files can be obtained from most program packages. 

Therefore, we have implemented a \texttt{DensityEvaluator} module that can process \textsc{Molden} files generated by
quantum-chemical program packages using Gaussian-type orbitals. This module makes use of \textsc{PySCF} to read in the basis
functions, molecular orbital coefficients, as well as occupation numbers and uses these to calculate the electron density,
its derivatives, and the corresponding electrostatic potential on the points of an integration grid. These quantities are 
provided as \texttt{GridFunction} objects inside \textsc{PyEmbed}. We note that the use of such a non-standardized legacy file format 
can cause problems, in particular because different program packages implement the \textsc{Molden} file format inconsistently. 
Further details are described in Ref.~\citenum{focke_coupled-cluster_2023}.

Currently, \textsc{PyEmbed} supports the generation of electron densities and related quantities via the \texttt{DensityEvaluator} 
\textsc{Molden} file interface from \textsc{Dalton}, \textsc{NWChem}, \textsc{Orca}, and \textsc{Turbomole}. Extending this interface 
to modern, standardized file formats for the exchange of molecular orbital / wavefunction information\cite{q5cost, smith_molssi_2021,
trexio} will be straightforward once such interfaces are available in the respective quantum-chemical program packages.

\subsection{Calculation of Embedding Potential}

At the core of \textsc{PyEmbed} is the \texttt{EmbPot} module, which performs the calculation of the embedding potential as given
in Eq.~\eqref{eq:embpot}. In this module, the nuclear potential of the frozen subsystems is evaluated directly from the corresponding 
atomic coordinates, whereas the Coulomb potential of the frozen subsystems needs to be provided as input as a suitable
\texttt{GridFunction} object. For evaluating the nonadditive kinetic and exchange--correlation potentials, it takes the electron 
density (and possibly its first and second derivatives) of the active and frozen subsystems as input, again in the form of 
suitable \texttt{GridFunction} objects. To this end, the functional derivatives of the selected approximate functionals are evaluated 
with the help of the \textsc{XcFun} library\cite{xcfun, xcfun-2-2-1}. For the nonadditive xc and kinetic energy functionals, both
LDA functionals (requiring only the electron densities of the subsystems) as well as GGA functionals (additionally requiring 
the first and second derivatives of the subsystem electron densities) are currently supported. Note that for GGA functionals,
the local embedding potential is evaluated explicitly, such that the matrix elements of the embedding potential can be evaluated
directly according to Eq.~\eqref{eq:ao-emb-ints-numint} (i.e., the derivatives of the basis functions are not needed).

As a result, this provides a \texttt{GridFunction} representation of the embedding potential, which can be imported into the
quantum-chemical calculation for the active subsystem. The evaluation of the matrix elements of the embedding potential
[see Eq.~\eqref{eq:ao-emb-ints}] by numerical integration generally needs to be performed inside the respective codes,
which also need to provide an interface for importing this embedding potential from file. Such interfaces are currently
publicly available in \textsc{Ams}/\textsc{Adf}, \textsc{Dalton}, \textsc{Dirac}, \textsc{NWChem},  \textsc{Molcas}, and 
\textsc{QuantumEspresso}.
We note that with the modular approach described here, the integration grid used inside the quantum-chemical program
packages in the case of DFT calculations and those used for the evaluation of the embedding potential and for the evaluation
of the matrix elements of the embedding potential according to Eq.~\eqref{eq:ao-emb-ints-numint} are generally different. 
Therefore, both grid have to be chosen appropriately.

\section{Applications}

\subsection{DFT-in-DFT and WFT-in-DFT embedding}
\label{sec:wftindft}

The exchange of grid-based data has been extensively employed by some of us in applications of WFT-in-DFT, employing a linearized embedding potential. In the case one is interested in energies for ground or excited states, the linearized scheme has been found to be rather accurate, since electron densities from WFT and DFT are often very similar\cite{gomes_calculation_2008,gomes_towards_2013,hfener_molecular_2012}. 

\begin{figure}[!h]
\centering
\includegraphics[width=1.0\textwidth]{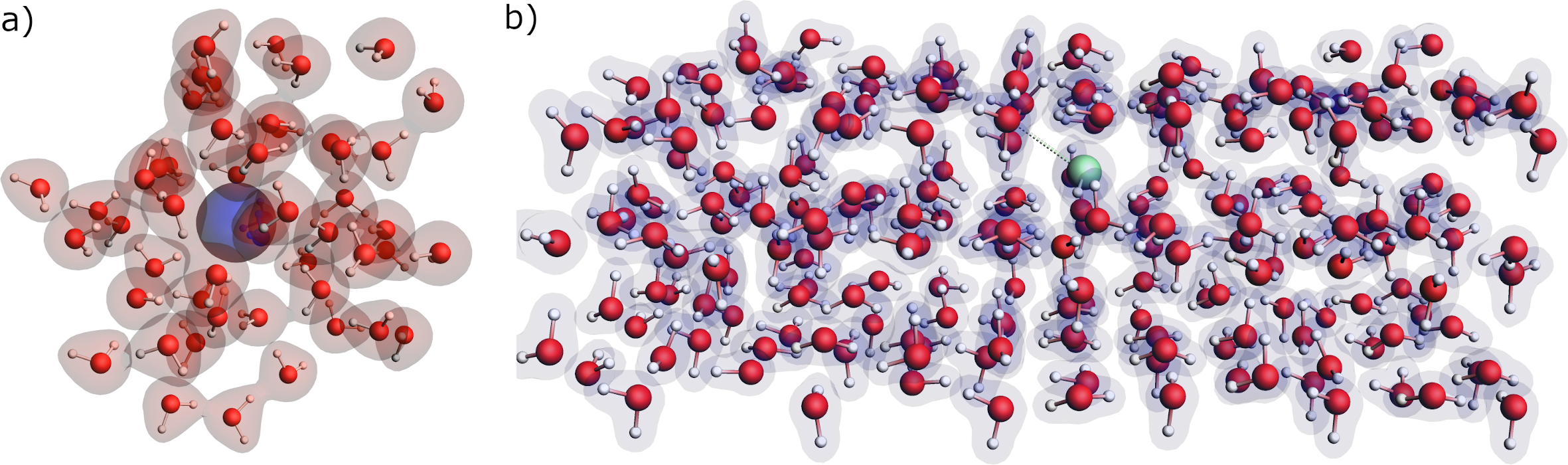}
\caption{\linespread{1.0}\selectfont
              Structural models of (a) a halide in a 50-water droplet used in the WFT-in-DFT calculations reported in 
              Ref.~\citenum{bouchafra_predictive_2018} (Figure available from 
              \url{https://doi.org/10.5281/zenodo.1477102} under a CC-BY license); and
              (b) HCl adsorbed onto an amorphous ice surface used in the WFT-in-DFT calculations 
              reported in Ref.~\citenum{opoku_simulating_2022} (Figure available from 
              \url{https://zenodo.org/doi/10.5281/zenodo.10591687} under a CC-BY license). In both cases the highlighted volumes indicate the electron densities for the individual subsystems, obtained from the preparatory DFT-in-DFT calculations.}
\label{fig:halides-surfacse-ice-hcl}
\end{figure}

One recent example is found in the simulations of valence binding energies of halides in water droplets\cite{bouchafra_predictive_2018}, such as the one shown in Fig.~\ref{fig:halides-surfacse-ice-hcl}a. In this work, the halide binding energies have been calculated with \textsc{Dirac} using the relativistic equation of motion coupled cluster for ionization energies (EOM-IP), with the embedding potential being obtained after preparatory DFT-in-DFT calculations with \textsc{ADF}. Due to the need to account for finite temperature effects, these CC-in-DFT calculations have been carried out for a set of \num{100}~snapshots originating from classical molecular dynamics simulations with polarizable force fields. With this setup, the overall computational cost for the calculation of each snapshot was roughly equivalent to that of the DFT-in-DFT calculations (instead of the much more computationally expensive EOM-IP calculation), while results for the halide binding energies were found to deviate from experiment by about \qty{0.1}{\electronvolt}, a value comparable to that obtained with state of the art periodic Green's functions calculations\cite{gaiduk_photoelectron_2016}.

A similar protocol, based on an ensemble of preparatory DFT-in-DFT calculations followed by a WFT-in-DFT calculation, has been employed to obtain the core electron binding energies associated with the chlorine 2s and 2p levels for chloride in water droplets as well as chloride and HCl on ice surfaces (see Fig.~\ref{fig:halides-surfacse-ice-hcl}b)\cite{opoku_simulating_2022}. The WFT-in-DFT calculations were again carried out with \textsc{Dirac}, but now employing the core-valence separation (CVS) approximation to the EOM-IP problem. As with the valence electron binding energies, for chloride these simulations compared quite favorably to experiment in both environments. For HCl, on the other hand, preliminary findings point to the need to improve the structural models to accurately capture  the predissociation tendency of HCl in contact with ice surfaces. 

More recently, DFT-in-DFT calculations combining the \textsc{Adf} code and the four-component code \textsc{Bertha} \cite{belpassi_bertha:_2020} through the use of the \textsc{PyBertha}\cite{de_santis_pyberthart:_2020} framework (which in turn employs \textsc{PyEmbed} as a backend for communicating with \textsc{Adf}) have been carried out to investigate the performance of FDE in describing the effects of solvation of gold clusters [for species such as \ce{Au8(H2O)_{10}} or \ce{Au4(H2O)_{80}}] and of confinement on superheavy elements (Rn, Cn, Fl, Og) encapsulated in \ce{C_{60}} molecules\cite{de_santis_frozen-density_2022}. For the case of superheavy elements, it was found that embedding potentials were largely transferable between atoms, and provided a more physical description of the cage than the previously employed potentials consisting of a square well. A key implementation detail in \textsc{PyBertha} for achieving high efficiency is the expression of the embedding potential on the basis of fit functions, instead of orbital pairs\cite{de_santis_frozen-density_2022}.

The applications outlined above also serve to underscore that when different subsystems are treated with different Hamiltonians, exchanging grid-based data has a practical (if not fundamental) advantage over data exchange through orbital manipulations: since water molecules or \ce{C_{60}} are made up of light elements for which spin-orbit coupling is not important, preparatory DFT-in-DFT calculations could safely be carried out with the scalar relativistic ZORA Hamiltonian (decreasing the overall cost of calculations), and the resulting embedding potential imported into four-component calculations without any particular difficulty. We are not aware of any other implementation that allows for such a mixture.

Grid-based data exchange has also been deployed for relativistic DFT-in-DFT calculations of response properties. In a recent application, the X-ray absorption spectra of the uranyl tetrachloride molecule (\ce{Cs2UO2Cl4}) has been obtained from the calculation of dipole polarizabilities within the damped response formalism\cite{misael_core_2023}, and in line with what had been found for valence excitations\cite{gomes_towards_2013}, embedding calculations in which the chlorides are represented by an embedding potential can reproduce very well the spectral features for the uranium $\mathrm{M_4}$ and $\mathrm{L_3}$ edges, and the O~K-edge.

Another example of DFT-in-DFT response calculations is that of  magnetic properties such as NMR shieldings, indirect spin-spin coupling constants, and magnetizabilities\cite{olejniczak_calculation_2017}. In this case, the embedding was determined self-consistently at the four-component level, followed by that of perturbed densities, required for the calculation of kernel and coupling contributions which are essential for linear response properties. 

It should be noted that the Fortran implementation in \textsc{Dirac} closely followed the design philosophy of \textsc{PyEmbed}, with the definition of \texttt{GridFunction} containing grid-based data for particular subsystems, and operations that can be carried out on them. This makes for a natural connection to the \textsc{PyAdf} and \textsc{PyEmbed} frameworks, and easily allows the exchange between subsystems (as \textsc{Dirac} is not capable of handling multiple fragments in the same execution). 

The magnetic properties embedding code has been used to investigate the solvent effects on NMR shieldings of the molybdate (\ce{MnO4^{2-}}) ion\cite{halbert_investigating_2020}, with temperature effects being taken into account by averaging the property calculations over \num{100}~snapshots from AIMD simulations of molybdate with \num{20}~water molecules\cite{nguyen_evaluation_2015}. One key finding from these studies is that despite the operators involved in the magnetic perturbations exhibiting an effective $r^{-2}$ dependence (suggesting strong locality), the response properties still show non-negligible contributions from longer-range interactions with solvent molecules. These findings, along with the challenges in understanding the interplay between long-range and short-range contributions, have motivated some of us to pursue the development of real-space data analysis and visualization, discussed below.

\subsection{Real-Time TDDFT-in-DFT Simulations}
\label{sec:rt}

The applications of FDE discussed above for calculating molecular properties rely in one way or another on perturbation (response) theory. Apart from the fact that response theory will break down in the case of strong external perturbations (e.g., in the simulation of processes such as high harmonic generation), and that they are unable to provide detained information in the attosecond regime, a practical drawback of response theory is the need to calculate first or higher-order derivatives of the embedding potential. 

An alternative approach lies in the so-called real-time (rt) methods, in which the time-dependent Schr\"odinger (or Dirac) equation is solved explicitly. In the case of Kohn-Sham DFT, this translates into
\begin{equation}
    \label{tdkseq}
    {\rm i}\frac{\partial \psi_{i}(\boldr,t)}{\partial t} = \left( -\frac{\nabla_{i}^2}{2} + v^\text{KS}_{\mathrm{eff}}(\boldr,t) 
       + v_{\rm emb}(\boldr, t) \right) \psi_{i}(\boldr,t)
\end{equation}
where from time-dependent orbitals $\psi_{i}(\boldr,t)$ one obtains the time-dependent density matrix $\textbf{D}_{\mu\nu}(t)$ and consequently the time-dependent electron density can be expressed as $\rho(\boldr,t) = \sum_{\mu\nu}\textbf{D}_{\mu\nu}(t) \chi_\mu(\boldr)\chi_\nu(\boldr)$, with $\chi_\mu(\boldr)$ being, for instance, atom-centered basis functions. This equation can be recast as the \textit{Liouville-von Neumann} (LvN) equation\cite{jakowski_liouvillevon_2009,li_time-dependent_2005}
\begin{align}
        i\frac{\partial \textbf{D}(t)}{\partial t} = \textbf{F}(t) \textbf{D}(t) -  \textbf{D}(t) \textbf{F}(t),
\end{align}
with the time-dependent Fock matrix $\textbf{F}(t)$ now includes the embedding potential. Molecular properties can then be obtained from a simulation of the time evolution of the dipole moment or another suitable operator, and this signal can be subsequently Fourier transformed to obtain for instance an excitation spectrum. 

One consequence of this formulation for the density-based embedding theories~\cite{krishtal_subsystem_2015,de_santis_environmental_2020}, is that the embedding potential now formally becomes time-dependent, even if only subsystem $I$ evolve in time (in analogy to the time-independent FDE case), due to the dependence of the embedding potential on both electron densities through the non-additive term (in what follows we only discuss results employing non-relativistic Hamiltonians, but note that a four-component code for rt-TDDFT-in-DFT calculations based on the \textsc{PyBertha} framework is currently under development). 

In practice, recalculating the embedding potential at each time step, especially considering that typical simulations require thousands to tens of thousands of time steps, can be computationally very expensive and, in some cases, prohibitive. However, in a first implementation with atom-centered Gaussian basis sets employing the \textsc{PyEmbed} framework and combining the \textsc{Adf} and \textsc{Psi4} codes, some of us have shown~\cite{de_santis_environmental_2020} that it is nevertheless possible to avoid recalculating the embedding potential over a certain (problem-dependent) number of time steps, and with that reduce the computational cost of rt-TDDFT-in-DFT FDE calculations, making it feasible to investigate the effect of the solvent on spectra with it.  

In spite of the computational savings introduced in Ref.~\citenum{de_santis_environmental_2020}, calculations with relatively large frozen density region remained rather time-consuming, putting application to systems requiring  large solvation shells (such as the halides discussed above) out of reach. At the same time, even for halides FDE had been shown to fail to properly describe the spectra of species such as fluoride\cite{bouchafra_predictive_2018}, which interact strongly with their first solvation shell.

In order to overcome these two difficulties, we have recently introduced\cite{martinez_b_solvation_2024} an embedding rt-TDDFT implementation based on the \textsc{PyEmbed} and \textsc{PyBertha-RT}\cite{de_santis_pyberthart:_2020} frameworks that combines the real-time\cite{koh_accelerating_2017,de_santis_environment_2022} Block-Orthogonalized Manby-Miller Embedding (BOMME)\cite{ding_embedded_2017} and FDE~\cite{de_santis_environmental_2020} approaches. This hybrid approach aims to reduce computational cost further while accurately treating strong interactions in the vicinity of a species of interest (BOMME). Moreover, it can incorporate long-range interactions (FDE), which are crucial for describing excitation processes accurately.

We also explored a simple way to speed up calculations by taking advantage of the GPU offloading capabilities of optimized tensor libraries like \textsc{PyTorch} and \textsc{Tensorflow}. Our early stage implementation, described in Ref.~\citenum{martinez_b_solvation_2024}, involved replacing certain \textsc{Numpy} operations, such as Fock matrix formation in time propagation, and constructing the matrix representation of the embedding potential, which are all performed in our own Python codes. Further work still needs to be done to fully realize the potential of GPU acceleration, in particular by enabling GPU offloading in more components of our implementation, as the generation of the non-additive potentials (\texttt{get\_nadd\_pot}) with \textsc{PyEmbed}. 

\begin{figure}[!h]
\begin{center}
  \includegraphics[width=1.0\textwidth]{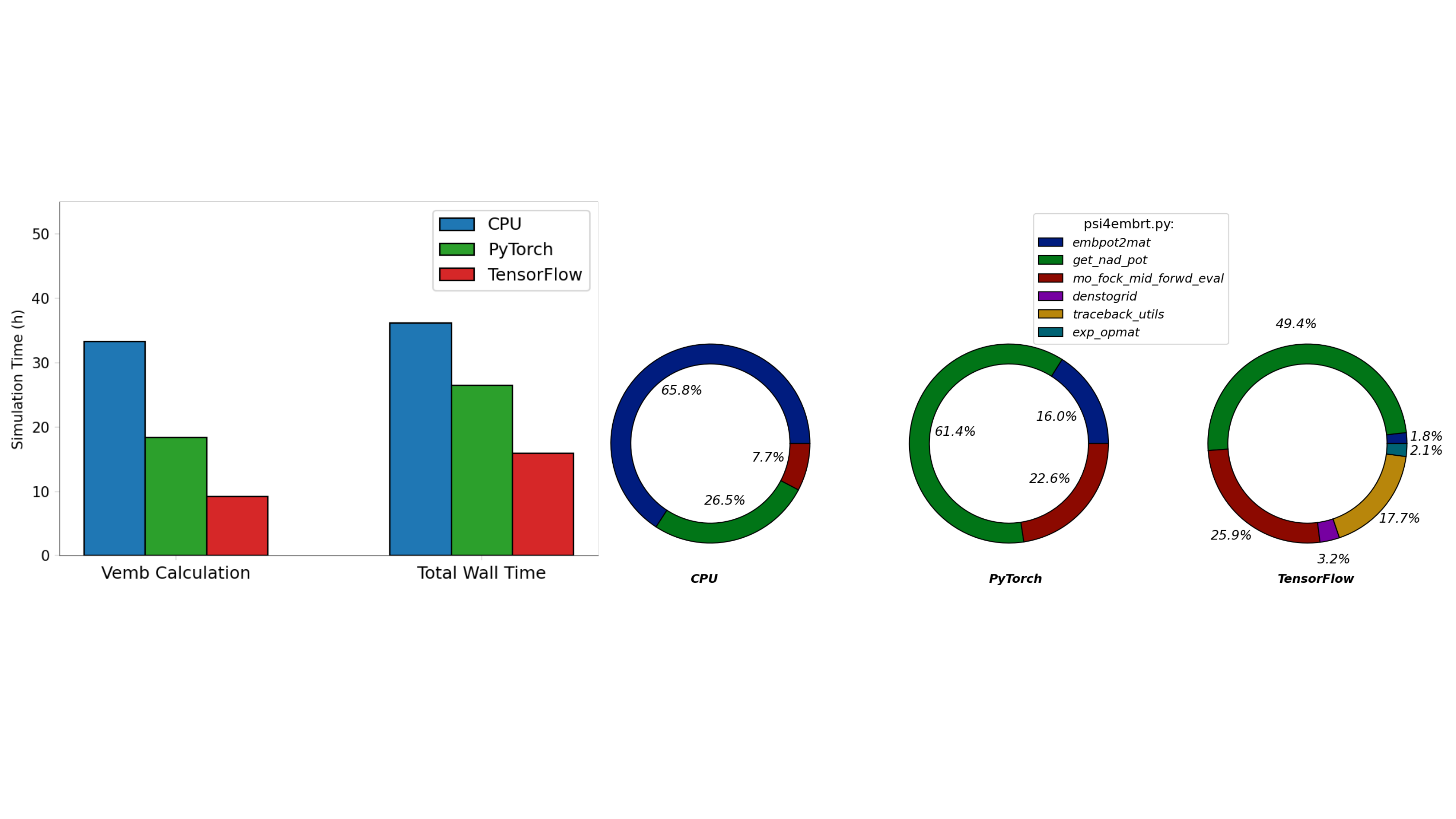}
\end{center}
\caption{\linespread{1.0}\selectfont
              Comparison of the performance of rt-TDDFT-in-DFT simulations with and without GPU offloading employing the 
              \textsc{PyTorch} and \textsc{Tensorflow} frameworks. Left: comparison of total simulation time (wall time) required to carry out 56000 time steps, and the aggregate of time spent 
              on evaluating the embedding potential. Right: breakdown of the time spent on the functions: \texttt{embpot2mat} 
              calculates matrix elements of the embedding potential in the electronic structure code; \texttt{get\_nad\_pot} calculates 
              the embedding potential on the grid with \textsc{PyEmbed}, \texttt{mo\_fock\_mid\_forward} extrapolates of Fock 
              matrices during the propagation; \texttt{denstogrid} calculates the electron density on the grid. 
              See Ref.~\citenum{martinez_b_solvation_2024} for a detailed discussion. The benchmarks were carried out using 64 OpenMP threads on an Intel(R) Xeon(R) Gold 6230R CPU @ 2.10GHz for the CPU runs, and 1 OpenMP thread (on the same type of CPU) and 1 NVIDIA A100-PCIE-40GB GPU for the GPU runs (Figure available from 
              \url{https://zenodo.org/doi/10.5281/zenodo.10591687} under a CC-BY license).}
\label{fig:rt_perf}
\end{figure}

The performance analysis summarized in  Fig.~\ref{fig:rt_perf}, illustrates the impact of GPU offloading. In this figure we can see that the total wall time is reduced when going from the CPU-only implementation to that with \textsc{PyTorch}, with a further reduction with \textsc{Tensorflow}. This is mostly due to a significant reduction in the portion of the total wall time spent on building the atomic orbital (AO) matrix representation of the embedding potential (\texttt{embpot2mat}), from 65.8\% in the original CPU-based implementation (making it the bottleneck) to  22.6\% for \textsc{PyTorch} and further down to 1.8\% for \textsc{Tensorflow}. With these speedups and the lack of GPU offloading, \texttt{get\_nadd\_pot} then becomes the bottleneck for the tensor libraries, representing 61.4\% of walltime for \textsc{PyTorch} and  49.4\%  for \textsc{Tensorflow}. The percentage of time spent on Fock matrix formation (\texttt{mo\_fock\_mid\_forward}) also becomes a bit more important for the tensor libraries, though they are rather similar (22.6\% of  walltime for \textsc{PyTorch} and  25.9\%  for \textsc{Tensorflow}). For \textsc{Tensorflow} other routines such as \texttt{traceback\_utils} start to become important for the timings, though at this stage we consider them as artifacts of profiling runs, and we intend to explore this matter further once the modifications to \texttt{get\_nadd\_pot} are carried out. 

Taken together, these results provide a good first indication of the potential to speed up python-based scientific code by offloading costly operations onto GPU-aware frameworks, with minimal code changes in the scientific application. 

\begin{figure}[!h]
\begin{center}
  \includegraphics[width=1.0\linewidth]{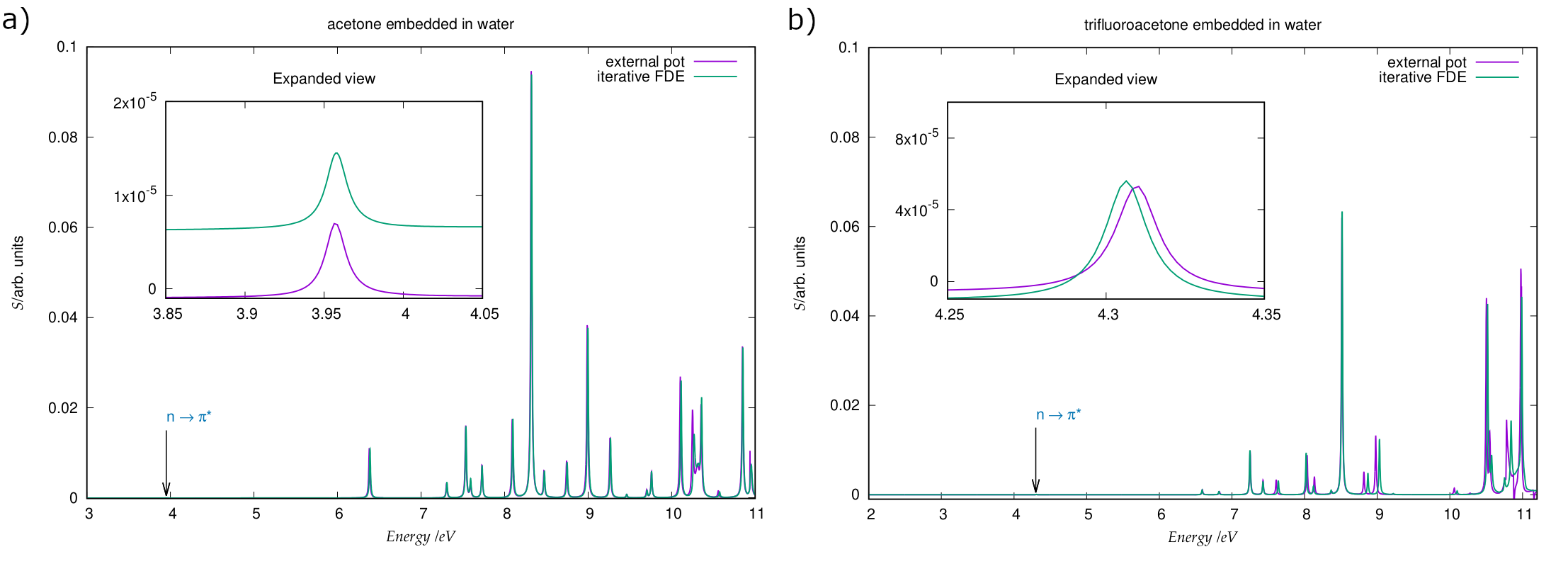}
\end{center}
\caption{\linespread{1.0}\selectfont
             Absorption spectra  (a) of acetone in water and obtained using the time-dependent embedding potential (green) evaluated 
              at each time step (denoted as iterative FDE) with the same simulation setup as in Ref.~\citenum{de_santis_environmental_2020}. 
              The violet trace (``external pot'') has been obtained here for simulations that closely match those in Ref.~\citenum{de_santis_environmental_2020}, 
              with the difference that there is no update of the embedding potential using the time propagation; and
              (b) of trifluoroacetone in water, embedded in external static potentials. We show in violet the spectrum obtained 
              with a potential originating from a calculation of acetone in water, and in green the spectrum obtained with the with an embedding 
              potential determined for trifluoroacetone in water.}
\label{fig:ace_water}
\end{figure}

In parallel to these developments, we have recently started to explore how far one can take the idea of not updating the embedding potentials at each timestep. To this end, in Fig.~\ref{fig:ace_water} we present results of a simulation revisiting the  acetone in water system of Ref.~\citenum{de_santis_environmental_2020} in which we compare two limiting cases: the update of the embedding potential at each timestep as done previously, and the opposite situation in which the embedding potential is not updated at all during the time propagation. These preliminary results seem to indicate that for this system, which is a case of relatively weak interactions between the active system and its environment, this very crude approximation yields results which are quite close to the original ones. This point requires further investigations, as it may be that this approximation will be suitable for a few low-lying transitions but break down for high energy transitions such as those in the X-ray region.

In the case of heavy atoms (Rn, Cn, Fl, Og) encapsulated in C$_{60}$, discussed above for time-independent calculations, we observed the transferability of the embedding potential. We explored here whether that could hold true also for time-dependent case, by considering trifluoracetone, which can be regarded as a modified acetone. We calculated its absorption spectrum utilizing embedding potentials without updates (corresponding to the assumption of weak interactions with the environment for the fluorinated species), one obtained for the acetone in water system, and another calculated for the trifluoracetone in water system. As shown in Fig.~\ref{fig:ace_water}b, the spectra for the two calculations show a semi-quantitative agreement. 
The conditions of transferability of embedding potentials in the case of confined molecular systems will be addressed in future works.  

\subsection{Density-Based Many-Body Expansion}
\label{sec:dbmbe}

Another example that has been enabled by the possibility to import and process grid-based quantities 
inside \textsc{PyEmbed} is the development of the density-based many-body expansion (db-MBE) by 
some of us \cite{schmitt-monreal_frozen-density_2020, schmitt-monreal_density-based_2021, 
focke_coupled-cluster_2023}. The db-MBE offers an accurate and efficient fragmentation method 
for the quantum-chemical treatment of molecular clusters. Its starting point is the conventional
many-body expansion (MBE) \cite{cui_theoretical_2006, richard_understanding_2014, herbert_fantasy_2019}, 
in which the total energy of a molecular cluster is approximated as [energy-based many-body expansion 
(eb-MBE)], 
\begin{equation}
  \label{eq:eb-mbe}
  E_\text{tot} \approx E_\text{eb-MBE}^{(n)} = 
      \sum_I E^{(1)}_I + \sum_{I < J} \Delta E^{(2)}_{IJ} + \dotsb + \sum \Delta E^{(n)}_{IJK\dotsm},
\end{equation}
where $E^{(1)}_I$ is the energy of the $I$-th fragment, $\Delta E^{(2)}_{IJ} = E^{(2)}_{IJ} - E^{(1)}_I - E^{(1)}_J$ 
is the interaction energy for the dimer of fragments $I$ and $J$, and so forth.

The db-MBE seeks to improve upon the eb-MBE by calculating the density-based correction,
\begin{equation}
  E_\text{db-corr}^{(n)} = E_\text{tot} \Big[ \rho^{(n)}_\text{tot}(\boldsymbol{r}) \Big] - E_\text{tot}^{(n)},
\end{equation}
where 
\begin{equation}
  \label{eq:mbe-density}
  \rho_\text{tot}(\boldsymbol{r}) \approx \rho_\text{tot}^{(n)}(\boldsymbol{r}) 
        =  \sum_I \rho^{(1)}_I(\boldsymbol{r}) + \sum_{I < J} \Delta \rho^{(2)}_{IJ}(\boldsymbol{r})
            + \dotsb + \sum \Delta \rho^{(n)}_{IJK\dotsm}(\boldsymbol{r}),
\end{equation}
is the many-body expansion of the total electron density, $E_\text{tot}[\rho]$ is the DFT total energy
functional, and $E_\text{tot}^{(n)}$ is its many-body expansion. With this density-based correction,
an improved $n$-body approximation to the total energy of the molecular cluster is obtained as
$E_\text{db-MBE}^{(n)} = E_\text{eb-MBE}^{(n)} + E_\text{db-corr}^{(n)}$. Previous studies have demonstrated the excellent accuracy of db-MBE in predicting both absolute and relative energies of water and ion-water clusters, even at the two-body expansion level \cite{schmitt-monreal_density-based_2021, schrmann_accurate_2023}.

At first order (i.e., using only the electron densities of the monomers), the density-based correction
can be calculated as
\begin{align}
  \label{eq:dbcorr-1}
  E^{(1)}_\text{corr} =& \sum_{I\ne J} \int \rho^{(1)}_I(\boldsymbol{r}) v^{(J)}_\text{nuc}(\boldsymbol{r}) \, {\rm d}^3r
                                    + \sum_{I \neq J} \frac{1}{2} \int \rho^{(1)}_I(\boldsymbol{r}) v_\text{Coul}[\rho^{(1)}_J](\boldsymbol{r})\, {\rm d}^3r             
  \nonumber \\
                               &+ E_\text{xc}^\text{nadd}\big[\{\rho^{(1)}_I\}\big] + T_s^\text{nadd}\big[\{\rho^{(1)}_I\}\big] + \sum_{I<J} E^{(IJ)}_\text{NN} ,
\end{align}
where the non-additive kinetic energy and exchange--correlation functionals
\begin{align}
  \label{eq:ts-nadd}
  T_s^\text{nadd}\big[\{\rho^{(1)}_I\}\big] 
     &= T_s\big[ \textstyle\sum_I \rho^{(1)}_I \big] - \sum_I T_s\big[\rho^{(1)}_I\big] \\
  \label{eq:exc-nadd}
  E_\text{xc}^\text{nadd}\big[\{\rho^{(1)}_I\}\big] 
     &= E_\text{xc}\big[ \textstyle\sum_I \rho^{(1)}_I \big] - \sum_I E_\text{xc}\big[\rho^{(1)}_I\big],
\end{align}
are evaluated using suitable approximate density functionals, and where $E^{(IJ)}_\text{NN}$ is 
the interaction energy between the nuclei of fragments $I$ and $J$. 

\begin{figure}[!h]
\begin{center}
  \includegraphics[width=0.6\textwidth]{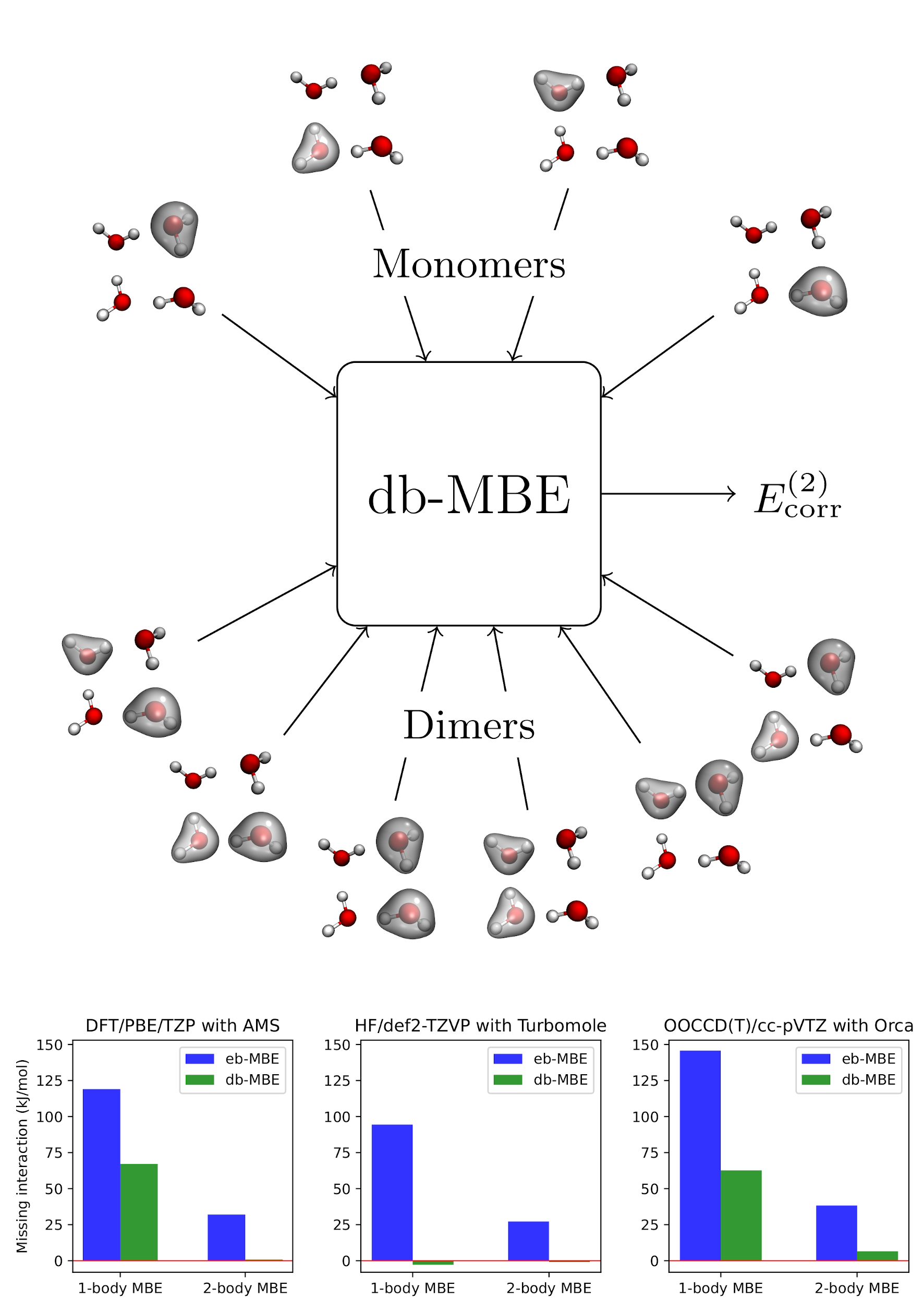}
\end{center}
\caption{\linespread{1.0}\selectfont
              Top: Illustration of the db-MBE for the example of a water tetramer. The electron densities and related quantities
              are fed into the db-MBE module of \textsc{PyAdf} as grid-based data. The db-MBE calculation uses these to
              calculate the density-based correction energy $E^{(2)}_\text{corr}$. Bottom: Error in interaction energy obtained 
              for the eb-MBE and db-MBE at first and second order, using fragment calculations with three different 
              quantum-chemical methods and program packages. The calculations have been performed as described 
              in Ref.~\citenum{focke_coupled-cluster_2023}.}
\label{fig:dbmbe}
\end{figure}

With representations of the monomers' electron densities (as well as their derivatives) and nuclear
and Coulomb potentials available, all terms in Eq.~\eqref{eq:dbcorr-1} can be evaluated using numerical
integration \cite{schmitt-monreal_frozen-density_2020}. We have integrated a db-MBE module in
\textsc{PyAdf} that makes use of the \textsc{PyEmbed} module as described above and evaluates
the density-based correction independent of external quantum-chemical program packages (see 
top part of Fig.~\ref{fig:dbmbe}). Again, the density functionals are evaluated using the \textsc{XcFun} 
library\cite{xcfun, xcfun-2-2-1}.

For higher orders ($n>1$), the density-based correction is given by
\begin{align}
  \label{eq:dbcorr-2}
  E^{(n)}_\text{corr} 
     &= \Big( V_\text{nuc}\big[ \rho_\text{tot}^{(n)} \big] - V_\text{nuc}^{(n)} \Big)
         + \Big( J\big[ \rho_\text{tot}^{(n)} \big] - J^{(n)} \Big) 
     \nonumber \\
     &\hspace{1cm}
         + T_s^{\text{nadd},(n)}\big[ \{\rho_I\}, \{\rho_{IJ}\}, \dotsc \big] + E_\text{xc}^{\text{nadd},(n)}\big[ \{\rho_I\}, \{\rho_{IJ}\}, \dotsc \big],
\end{align}
which is also evaluated within the db-MBE module of \textsc{PyAdf} using the corresponding $n$-mer densities and potentials. 

The nuclear attraction and Coulomb energy functionals, $V_\text{nuc}[\rho]$ and $J[\rho]$ are calculated for the 
$n$-body expansion of the electron density [cf.~Eq.~\eqref{eq:mbe-density}] using numerical integration, where 
the Coulomb energy is evaluated by integrating the electron density with the corresponding Coulomb potential. 
To this end, the many-body expansions of the electron density and of the Coulomb potential are assembled inside 
the db-MBE module of \textsc{PyAdf} from the grid-based representations of the individual $n$-mer densities and 
potentials.

In Eq.~\eqref{eq:dbcorr-2}, $V_\text{nuc}^{(n)}$ and $J^{(n)}$ refer to the many-body expansion of the nuclear
attraction and Coulomb energies, respectively, are defined in analogy to Eq.~\eqref{eq:eb-mbe}. The individual
energy contributions required for these terms can either be extracted from the underlying quantum-chemical
calculations for the $n$-mers, or can be evaluated by numerical integration in the db-MBE module from the 
electron densities and Coulomb potentials obtained from each of the $n$-mer calculation. Finally, the $n$-body 
nonadditive kinetic and exchange--correlation energy functionals, $T_s^{\text{nadd},(n)}$ and 
$E_\text{xc}^{\text{nadd},(n)}$ are defined in analogy to Eq.~\eqref{eq:ts-nadd} and~\eqref{eq:exc-nadd} (see
Ref.~\citenum{schmitt-monreal_frozen-density_2020} for details) and are evaluated using \textsc{XcFun}.

Our implementation of the db-MBE, facilitated by \textsc{PyAdf} and \textsc{PyEmbed}, exemplifies the
modular development of novel quantum-chemical fragmentation methods. The db-MBE requires the calculation 
of a density-based correction, which depends not only on energies from the quantum-chemical calculations for
the individual fragments, but also on the resultant electron densities. While such computations typically fall within 
the purview of quantum-chemical software packages, they can be realized within an external Python module through 
the grid-based exchange of electron densities and related quantities.

Given that the db-MBE module solely requires grid-based representations of the electron density and the Coulomb
potential as its input (besides the atomic coordinates of the fragments), it operates independently of the specific 
type of underlying quantum-chemical calculations. Therefore, the db-MBE can take densities generated 
using different quantum-chemical methods in different program packages (see bottom part of Fig.~\ref{fig:dbmbe}. 
This versatility has enabled the extension of the db-MBE from DFT calculations to coupled-cluster 
methods \cite{focke_coupled-cluster_2023}.

\subsection{Real-Space Data Analysis and Visualization}
\label{sec:gosia-ttk}

Another important area of application for workflows that involve the exchange of grid-based data is the analysis 
and visualization of quantum-chemical calculation. Such combined methods can be particularly insightful --- they 
adopt the ``give us insight (not) and numbers'' mindset \cite{coulson_present_1960, thiel_theoretical_2011, 
neese_chemistry_2019}, which has been driving the development of numerous analysis tools to accompany 
quantum-chemical calculations.


\begin{figure}
\centering
  \includegraphics[width=0.8\linewidth]{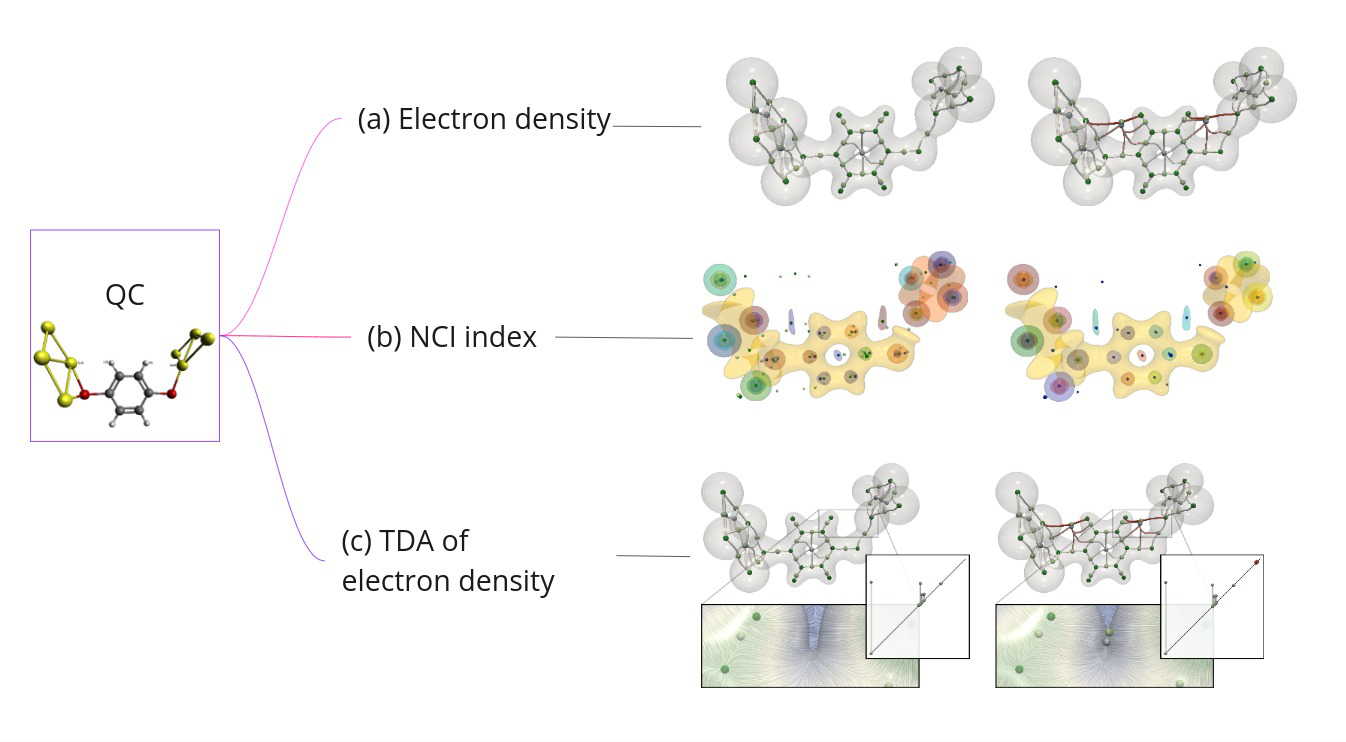}
  \caption{\linespread{1.0}\selectfont
               The selection of the correct molecular descriptor and the analysis method can significantly affect the 
               analysis results: The topological analysis of the electron density of the Au$_4$--S--C$_6$H$_4$--S'--Au'$_4$ 
               complex (a) by QTAIM and TDA suggests relativity-induced non-covalent interactions between Au and H atoms 
               (manifested by the absence/presence of the bond critical points in the non-relativistic/relativistic case, respectively, 
               demonstrated on the left and right figures). Their further inspection with the TDA tools (c) shows that these bond 
               critical points are insignificant, and their importance is further reduced with the improvement of the quantum 
               chemistry model. A better molecular descriptor for uncovering weak interactions is the NCI index (b), which 
               unambiguously points to the presence of the bond critical points irrespective of the relativity level applied in 
               QM calculations. Example after Ref.~\citenum{olejniczak_topological_2020}.}
\label{fig:visu-analysis}
\end{figure}

Specifically, we highlight techniques applied to \emph{molecular descriptors}, which can be represented as scalar, 
vector, or tensor fields and discretized on grids in physical space. Such molecular descriptors can be provided by 
quantum-chemical program packages by the interfaces described above (passed as \texttt{GridFunction}s, see Fig.~\ref{fig:visu-analysis}), and 
if necessary, the modules described in Section~\ref{sec:design} facilitate their processing, manipulation, and 
export to software tools for their analysis and manipulation.

The most obvious example of such a molecular descriptor is the electron density, $\rho(\boldr)$. Plotting the 
electron density in real space as, for instance, a three-dimensional surface for a selected isovalue, can demonstrate 
the ``shape'' of a molecule or a specific molecular fragment (depending on the isovalue), point to nuclear positions and 
identify chemical bonds (see Fig.~\ref{fig:visu-analysis}a). More elaborated electron density functions designed to capture 
specific information can highlight a particular feature of the molecular system. For instance, the electron density Laplacian 
$\Delta\rho(\boldr)$ identifies electron localization domains.\cite{contreras-garca_perspective:_2018}. The reduced density 
gradient, $s(\boldr) \propto |\nabla \rho (\boldr)|/\rho(\boldr)$, captures weak non-covalent interactions much better than the 
electron density. For this reason, its visualization (popularized as the NCI index\cite{johnson_revealing_2010, laplaza_nciplot_2021}) 
often complements electron density plots in studies on chemical interactions (see Fig.~\ref{fig:visu-analysis}b).
Such an information-centric approach to chemical simulations bloomed into the conceptual DFT theory\cite{geerlings_conceptual_2003} 
and Quantum Chemistry Topology (QCT)\cite{chauvin_applications_2016} domains, which largely focus on developing such molecular 
descriptors. 

Efficient visualization techniques for such fields can be broadly classified as geometrical or feature-based approaches (for more detailed classification schemes, see, e.g., Ref.~\citenum{lipa_visualization_2012}). Geometrical approaches are based on plotting the data either directly on the grids (as points or vectors) or by various 
functions that represent the data, for instance, via contours (scalar data) or integral curves (vector data). However, while 
these methods illustrate the underlying data intuitively and efficiently, they ``transfer the burden of feature recognition to 
human brains''\cite{wang_numerics_2016}.
Moreover, such methods come with various difficulties, which emerge especially for more complex data, such as the selection 
of meaningful isosurfaces (scalar data) or the choice of seeding points for streamlines (vector data), to name but a few. 
The other group of methods focuses on extracting specific features of the data, which are independent of the choice of their 
visual representation and, therefore, can lead to more reliable qualitative conclusions. An example is the topology-based 
visualization,\cite{heine_survey_2016} which, for instance, applied to a scalar field, results in the \emph{topological skeleton} 
of the field composed of the critical points and separatrices (lines and planes ``joining'' these points). 

Such topological methods have been widely adopted in computational chemistry. In particular, the whole QCT domain is based 
on the analysis of scalar and vector fields through their gradient fields. Specifically, it applies the theory of dynamical systems 
to the studied field gradient and interprets the extracted topological objects (e.g., critical points) in terms of chemical concepts 
(e.g., chemical bonds). The well-known example of a QCT method is the topological analysis of the electron density in molecular 
systems, pioneered by Bader and popularized as the Quantum Theory of Atoms in Molecules (QTAIM) \cite{bader_quantum_1991}.
However, the practical applications of QCT methods to chemistry problems are often burdened with technical difficulties, and their 
value is reduced by ambiguous interpretations. 

Modern analysis approaches, such as the Topological Data Analysis (TDA), can 
address many of such challenges. TDA is a new research field at the interface between mathematics and computer science, rooted 
in sound theoretical settings such as Morse theory\cite{milnor_morse_1963} and Persistent Homology (PH) \cite{edelsbrunner_computational_2009}. 
TDA constitutes an appealing framework for extracting the topological features in the data, measuring their salience (what allows 
for discriminating important features from non-essential ones), and for multi-scale data analysis (for examples from chemistry,
see Refs.~\citenum{xia_persistent_2018, wu_topps:_2018, pirashvili_improved_2018, hermosilla_physics-based_2017, 
gnther_characterizing_2014, bhatia_topoms:_2018}). As such, it is particularly useful to study the manifestation of weak phenomena in the data representing both weak and strong physical features. A good illustration of this aspect is the analysis of non-covalent interactions in small gold complexes, Au$_4$--S--C$_6$H$_4$--S'--Au'$_4$, based on the presence of bond critical points of the electron density in the region between interacting atoms: while QTAIM attributed their appearance to the relativistic effects,\cite{anderson_molecular_2019} TDA demonstrated that these critical points are non-essential electron density features\cite{olejniczak_topological_2020}. 

If the analysis outcome remains inconclusive, a good strategy is to reconsider the selection of a molecular descriptor for the studied phenomenon --- in the cited case, further studies employing the reduced density gradient scalar field confirmed the presence of non-covalent interactions between Au--H irrespective of the inclusion of relativistic effects in the quantum chemistry model (Fig.~\ref{fig:visu-analysis}c). The choice of an appropriate function for analysis is also crucial for understanding specific features of descriptors represented by higher-rank tensor fields. For example, the extraction of axial and toroidal vortices in the ``omega'' scalar field derived from the magnetically-induced current density tensor by TDA tools \cite{olejniczak_topological_2023} serves as a good illustration. While the primary focus of real-space analysis revolves around data visualization, other aspects such as quantification (e.g., through data integration), space partitioning, and data clustering can now also be efficiently addressed by TDA tools and libraries implementing them, such as the Topological Toolkit (TTK) \cite{ttk17,ttk19}.

\section{Conclusions and Outlook}
\label{sec:conclusion}

Realizing complex workflows in quantum chemistry that involve different quantum-chemical program packages presents a challenge, particularly in the context of quantum-chemical subsystem and embedding methods. Quantum chemistry software packages are typically developed as stand-alone codes that prioritize numerical efficiency, accuracy, and adaptability to parallel architectures, but they often lack standardized interfaces for interoperability. However, achieving interoperability requires the seamless exchange of complex data between different program packages.

In this context, we have introduced an approach implemented in the \textsc{PyEmbed} module of the \textsc{PyAdf} scripting framework to facilitate the exchange of electron densities, related properties, and embedding potentials between various quantum-chemical program packages. We have opted to utilize grid-based data and developed Python modules specifically for this purpose. Leveraging grid data for information exchange significantly simplifies the interfaces between program packages, as it is agnostic to internal data structures and technical intricacies such as basis functions and basis sets. The \textsc{PyEmbed} module outlined here establishes clear data and file formats for grid-based information exchange, streamlining the integration of program packages into workflows to the creation of clean interfaces that furnish the necessary grid-based data.

While grid-based data simplifies data exchange, standardization of data and file formats, such as adopting a standardized format based on HDF5, would greatly enhance the broader acceptance of the modular approach outlined here. Furthermore, establishing clear and standardized interfaces in quantum-chemical program packages for results data, including molecular orbital coefficients, will further improve interoperability. In the approach presented here, utilizing the \textsc{Molden} file format, despite its antiquated nature and potential limitations in meeting modern software development needs, offers the advantage of enabling the generation of grid-based data outside of quantum-chemical program packages, thus allowing this process to be performed within \textsc{PyEmbed}. This flexibility enhances the versatility of the modular approach and streamlines the integration of grid-based data into quantum-chemical workflows.

The Python modules discussed here are presently incorporated into the PyADF scripting framework but are also designed to be utilized independently. We intend to release \textsc{PyEmbed} as a standalone library in the near future. Additionally, we plan to introduce support for GPU offloading using machine learning frameworks such as \textsc{PyTorch} or \textsc{Tensorflow}. This integration has the potential to yield substantial speed improvements, particularly when processing grid-based data, aligning with our recent efforts to accelerate rt-TDDFT-in-DFT calculations \cite{martinez_b_solvation_2024}.

We have showcased numerous applications made possible by interoperable workflows that rely on the exchange of grid-based data between quantum-chemical program packages. Primarily, \textsc{PyEmbed} facilitates DFT-in-DFT and WFT-in-DFT embedding workflows by integrating various program packages, thereby granting access to a wide array of quantum-chemical methods. The modular nature of this approach is especially well-suited for implementing customized freeze-and-thaw schemes involving different program packages.

In a broader sense, \textsc{PyEmbed} streamlines the rapid prototyping of subsystem and embedding methods. This capability has been exemplified in the development of real-time time-dependent density functional theory (rt-TDDFT) embedding schemes, where different program packages must interact at each time step. Similarly, the creation of the density-based many-body expansion (db-MBE) has been made possible solely through the modular and interoperable approach outlined here. These instances signify a larger paradigm shift, wherein the innovation of novel quantum-chemical methods shifts away from monolithic codes to lightweight Python scripts that manipulate and amalgamate externally provided data. We anticipate that this evolving paradigm will gain broader acceptance in the years to come (see also Refs.~\citenum{jacob_open_2016, lehtola_call_2023}).

Finally, the utilization of grid-based data streamlines the incorporation of data analysis and visualization tools into quantum chemistry workflows. This facilitates the development of customized workflows that seamlessly integrate analysis tools to adjust the computational model as required, akin to the modeling cycle proposed in Ref.~\citenum{rommel_prescriptive_2021}. These adaptive workflows will need to encompass steps related to evaluating the accuracy of models and data, which are pivotal for their optimization and refinement. The integration of real-space data analysis could enhance numerous stages of such efficient modeling cycles.

The integration of grid-based data, including descriptors utilized in QSAR/QSAP\cite{karelson_quantum-chemical_1996} or from conceptual DFT\cite{geerlings_conceptual_2003}, into computational workflows also presents new opportunities for high-throughput applications. These applications can serve as the foundation for machine learning in computational chemistry\cite{dral_quantum_2020, dou_machine_2023}.

\section*{Data Availability}
\vspace{-2ex}

The most recent version of the \textsc{PyAdf} scripting framework, including the \textsc{PyEmbed} module are
the related functionality described in this article, is available at \url{https://github.com/chjacob-tubs/pyadf-releases/}.
Code and data for the applications reviewed here are available in the cited publications.

\section*{Author Contributions}
\vspace{-2ex}

\textbf{Kevin Focke}: Methodology (equal), Software (equal), Visualization (lead), 
  Writing – Original Draft (supporting), Writing -- Review and Editing (equal).
\textbf{Matteo De Santis}: Methodology (equal), Software (equal), Visualization (supporting), 
  Writing – Original Draft (supporting), Writing -- Review and Editing (equal).
\textbf{Mario Wolter}: Methodology (equal), Software (supporting), Supervision (supporting), 
  Writing -- Review and Editing (equal).
\textbf{Jessica Martinez}: Methodology (equal), Software (equal), Visualization (supporting), 
  Writing -- Review and Editing (equal).
\textbf{Valérie Vallet}: Conceptualization (supporting), Supervision (supporting), Methodology (supporting), 
  Writing -- Review and Editing (equal).
\textbf{André Severo Pereira Gomes}: Conceptualization (lead), Methodology (supporting), Software (equal), Supervision (equal), 
  Writing – Original Draft (lead), Writing -- Review and Editing (equal).
\textbf{Małgorzata Olejniczak}: Conceptualization (lead), Methodology (supporting), Software (equal), Supervision (equal), 
  Writing – Original Draft (lead), Writing -- Review and Editing (equal).
\textbf{Christoph R. Jacob}: Conceptualization (lead), Methodology (supporting), Software (equal), Supervision (equal), 
  Writing – Original Draft (lead), Writing -- Review and Editing (equal).

\section*{Conflicts of Interest}
\vspace{-2ex}

The authors have no conflicts to disclose.

\section*{Acknowledgments}

C.R.J. and A.S.P.G. gratefully acknowledge funding from the Franco--German project CompRIXS (Agence nationale de la 
recherche ANR-19-CE29-0019, Deutsche Forschungsgemeinschaft JA 2329/6-1). C.R.J. and M.W. acknowledge funding 
from the Deutsche Forschungsgemeinschaft for the development of \textsc{PyAdf} (Project Suresoft, JA 2329/7-1). M.D.S., 
V.V., and A.S.P.G acknowledge funding from projects CPER WaveTech, Labex CaPPA (Grant No. ANR-11-LABX-0005-01), 
the I-SITE ULNE project OVERSEE and MESONM International Associated Laboratory (LAI) (Grant No. ANR-16-IDEX-0004), 
as well support from the French national supercomputing facilities (Grant Nos. DARI A0130801859, A0110801859). 
J.A.M.B. acknowledges support from the Chateaubriand Fellowship of the Office for Science \& Technology of the 
Embassy of France in the United States, and the National Science Foundation under grants No. CHE-2154760 and OAC-2117429.
M.O. acknowledges support from the Polish National Science Centre (NCN) (grant number 2020/38/E/ST4/00614).

\linespread{1.1}\selectfont

\end{document}